\documentclass{emulateapj}
\shorttitle{The Eclipsing Overtone Cepheid 81.8997.87}
\slugcomment{Submitted to ApJ}
\begin{document}

\title{The Nature of the Companion to the Eclipsing Overtone Cepheid MACHO 81.8997.87}

\author{D.~Lepischak, D.L.~Welch}
\affil{Department of Physics and Astronomy, McMaster University\\
Hamilton, Ontario Canada L8S 4M1}
\email{lepischak, welch@physics.mcmaster.ca}
\and
\author{P.B.M.~van~Kooten}
\affil{Department of Physics, Queen's University\\
Kingston, Ontario Canada K7L 3N6}

\begin{abstract}

The lightcurve of the Large Magellanic Cloud (LMC) variable star MACHO
81.8997.87 shows evidence for photometric variations due to both
stellar pulsation, with a 2.035 day period, and eclipsing behavior,
with an 800.4 day period.  The primary star of the system has been
identified as a first-overtone Cepheid but the nature of the secondary star
has not been determined.  Here we present multicolor BVI photometry of
a primary eclipse of the system and fit a model to the complete
lightcurve to produce an updated set of elements.  These results are
combined with 2MASS JHK photometry to give further insight into the
identity of the companion star.  We find that the companion is most
consistent with a late-K or an early-M giant but also that there are a
number of problems with this interpretation.  The prospects for future
observations of this system are also discussed.

\end{abstract}


\section{Introduction}

There is a small but growing list of regularly pulsating stars that
are known to be members of eclipsing binary systems.  \cite{myfirst}
present observations and analysis of three eclipsing Cepheid variables
for which data exist in the MACHO Project Large Magellanic Cloud (LMC)
database.  \cite{RB01} list nine eclipsing binary systems that contain
$\delta$ Scuti variables and new candidates have been identified since
that time (see \cite{gkdra} and \cite{ecl_d}).  Most recently
\cite{eclRR} list three objects whose lightcurves show evidence for
eclipses and RR Lyr-type pulsations.  Although one or more of these
may be the result of photometric contamination, clearly this is a
burgeoning field for obtaining long-sought direct measurements of
pulsating star properties.

The astrophysical returns from systems that combine eclipsing and
pulsating behavior can be considerable.  An eclipsing Cepheid system,
if also a double-lined spectroscopic binary, can give a determination
of the mass and luminosity of the Cepheid that is not only more
accurate than existing measurements but also independent of assumed
distance estimates.  Such a system would offer an independent
calibration of the period-luminosity and period-luminosity-color
relations and the most direct measurement of the Cepheid's mass.

Here we present additional observations and an updated analysis of the
eclipsing Cepheid system MACHO 81.8997.87.  In particular, we more strongly 
constrain the nature of the system's secondary star. 

\section{Observations}

\begin{deluxetable}{ccc}
\small
\tablecolumns{3} 
\tablewidth{0pt}   
\tablecaption{B photometry of April 2001 primary eclipse obtained with
the 1.9 m telescope at MSO.\label{B-tab}}
\tablehead{
\colhead{HJD}&
\colhead{$B$}&
\colhead{$\sigma_B$}\\
&
\colhead{(mags)}&
\colhead{(mags)}
}
\startdata 
2451998.957875 & 18.219 & 0.030 \\
2451998.959796 & 18.298 & 0.013 \\
2451998.967262 & 18.306 & 0.014 \\
2451998.974738 & 18.198 & 0.024 \\
2452000.911406 & 18.342 & 0.013 \\
2452000.918872 & 18.338 & 0.013 \\
2452000.926349 & 18.347 & 0.013 \\
2452002.898562 & 18.379 & 0.016 \\
2452002.906039 & 18.435 & 0.019 \\
2452002.962833 & 18.360 & 0.020 \\
2452003.898742 & 18.306 & 0.015 \\
2452003.906219 & 18.363 & 0.016 \\
2452003.913696 & 18.314 & 0.016 \\
2452004.883170 & 18.528 & 0.018 \\
2452004.890646 & 18.512 & 0.018 \\
2452004.898112 & 18.601 & 0.018 \\
2452005.884403 & 18.445 & 0.020 \\
2452005.891869 & 18.400 & 0.018 \\
2452005.899358 & 18.414 & 0.018 \\
2452007.905100 & 18.598 & 0.020 \\
2452007.912577 & 18.665 & 0.023 \\
2452008.880999 & 18.773 & 0.020 \\
2452008.888475 & 18.804 & 0.022 \\
2452008.895952 & 18.823 & 0.021 \\
2452011.891229 & 18.443 & 0.013 \\
2452011.898694 & 18.458 & 0.016 \\
2452012.896051 & 18.598 & 0.014 \\
2452012.903528 & 18.591 & 0.014 \\
2452012.911004 & 18.589 & 0.012 \\
2452013.881219 & 18.228 & 0.012 \\
2452013.889228 & 18.226 & 0.012 \\
2452013.896705 & 18.241 & 0.013 \\
2452014.878934 & 18.434 & 0.016 \\
2452014.886411 & 18.440 & 0.012 \\
2452014.923345 & 18.451 & 0.013 \\
2452014.931516 & 18.438 & 0.012 \\
2452014.938981 & 18.463 & 0.012 \\
2452015.865342 & 18.060 & 0.014 \\
2452015.873236 & 18.092 & 0.014 \\
2452015.880713 & 18.095 & 0.013 \\
\enddata  
\end{deluxetable}

\begin{deluxetable}{ccc}
\small
\tablecolumns{3} 
\tablewidth{0pt}   
\tablecaption{V photometry of April 2001 primary eclipse obtained with
the 1.9 m telescope at MSO.\label{V-tab}}
\tablehead{
\colhead{HJD}&
\colhead{$V$}&
\colhead{$\sigma_V$}\\
&
\colhead{(mags)}&
\colhead{(mags)}
}
\startdata 
2451998.933593 & 17.208 & 0.037 \\
2451998.935768 & 17.175 & 0.037 \\
2451998.941289 & 17.168 & 0.036 \\
2451998.947667 & 17.181 & 0.035 \\
2452000.873177 & 17.192 & 0.038 \\
2452000.874856 & 17.118 & 0.038 \\
2452000.875909 & 17.234 & 0.038 \\
2452000.877610 & 17.231 & 0.038 \\
2452000.879288 & 17.221 & 0.036 \\
2452000.883281 & 17.228 & 0.036 \\
2452000.887286 & 17.227 & 0.036 \\
2452001.083492 & 17.426 & 0.036 \\
2452001.085402 & 17.414 & 0.037 \\
2452002.882451 & 17.256 & 0.036 \\
2452002.887138 & 17.246 & 0.036 \\
2452002.891837 & 17.238 & 0.035 \\
2452003.867689 & 17.224 & 0.039 \\
2452003.870363 & 17.115 & 0.038 \\
2452003.872330 & 17.151 & 0.037 \\
2452003.874402 & 17.165 & 0.035 \\
2452003.882411 & 17.162 & 0.035 \\
2452004.045387 & 17.237 & 0.042 \\
2452004.049391 & 17.257 & 0.046 \\
2452004.053384 & 17.226 & 0.042 \\
2452004.864304 & 17.383 & 0.039 \\
2452004.866561 & 17.376 & 0.040 \\
2452004.868818 & 17.359 & 0.036 \\
2452004.872834 & 17.388 & 0.036 \\
2452005.034086 & 17.316 & 0.038 \\
2452005.038091 & 17.355 & 0.037 \\
2452005.042095 & 17.364 & 0.043 \\
2452005.863281 & 17.345 & 0.040 \\
2452005.864751 & 17.303 & 0.040 \\
2452005.866336 & 17.329 & 0.038 \\
2452005.868315 & 17.308 & 0.040 \\
2452005.870306 & 17.313 & 0.036 \\
2452005.874311 & 17.332 & 0.037 \\
2452005.878315 & 17.315 & 0.037 \\
2452007.869857 & 17.427 & 0.039 \\
2452007.871211 & 17.468 & 0.040 \\
2452007.872357 & 17.442 & 0.037 \\
2452007.876350 & 17.443 & 0.038 \\
2452008.866519 & 17.613 & 0.039 \\
2452008.870512 & 17.621 & 0.039 \\
2452008.874517 & 17.605 & 0.039 \\
2452011.862849 & 17.323 & 0.037 \\
2452011.863856 & 17.333 & 0.037 \\
2452011.865071 & 17.320 & 0.037 \\
2452011.868046 & 17.322 & 0.037 \\
2452011.869736 & 17.310 & 0.035 \\
2452012.042792 & 17.313 & 0.036 \\
2452012.046785 & 17.252 & 0.036 \\
2452012.881861 & 17.427 & 0.040 \\
2452012.885854 & 17.417 & 0.038 \\
2452012.889859 & 17.419 & 0.038 \\
2452013.027788 & 17.404 & 0.038 \\
2452013.031781 & 17.372 & 0.038 \\
2452013.035774 & 17.369 & 0.037 \\
2452013.915965 & 17.155 & 0.035 \\
2452013.924136 & 17.164 & 0.035 \\
2452014.865566 & 17.291 & 0.038 \\
2452014.869559 & 17.292 & 0.036 \\
2452014.873552 & 17.291 & 0.037 \\
2452015.897461 & 17.028 & 0.035 \\
2452015.999127 & 17.052 & 0.035 \\
2452016.003132 & 17.057 & 0.035 \\
2452016.007125 & 17.061 & 0.036 \\
\enddata  
\end{deluxetable}

\begin{deluxetable}{ccc}
\small
\tablecolumns{3} 
\tablewidth{0pt}   
\tablecaption{I photometry of April 2001 primary eclipse obtained with
the 1.9 m telescope at MSO.\label{I-tab}}
\tablehead{
\colhead{HJD}&
\colhead{$I$}&
\colhead{$\sigma_I$}\\
&
\colhead{(mags)}&
\colhead{(mags)}
}
\startdata 
2451998.899275 & 15.721 & 0.017 \\
2451998.902181 & 15.694 & 0.022 \\
2451998.904171 & 15.707 & 0.018 \\
2451998.905977 & 15.718 & 0.022 \\
2451998.909171 & 15.735 & 0.022 \\
2451998.913697 & 15.720 & 0.024 \\
2452000.903548 & 15.723 & 0.010 \\
2452000.905469 & 15.720 & 0.011 \\
2452000.907379 & 15.716 & 0.012 \\
2452002.933307 & 15.735 & 0.011 \\
2452002.935564 & 15.729 & 0.010 \\
2452002.937833 & 15.727 & 0.011 \\
2452003.923349 & 15.727 & 0.010 \\
2452003.925617 & 15.726 & 0.010 \\
2452003.927874 & 15.720 & 0.010 \\
2452004.069056 & 15.765 & 0.015 \\
2452004.071324 & 15.762 & 0.016 \\
2452004.073593 & 15.769 & 0.016 \\
2452004.917476 & 15.851 & 0.010 \\
2452004.919732 & 15.855 & 0.010 \\
2452004.922002 & 15.861 & 0.010 \\
2452005.058588 & 15.850 & 0.015 \\
2452005.060857 & 15.844 & 0.016 \\
2452005.063125 & 15.849 & 0.015 \\
2452005.918062 & 15.837 & 0.013 \\
2452005.920330 & 15.849 & 0.014 \\
2452005.922599 & 15.838 & 0.014 \\
2452006.070412 & 15.863 & 0.010 \\
2452006.072680 & 15.877 & 0.011 \\
2452006.074948 & 15.851 & 0.012 \\
2452007.896258 & 15.994 & 0.029 \\
2452007.898515 & 15.979 & 0.023 \\
2452007.900783 & 16.002 & 0.027 \\
2452008.916242 & 16.017 & 0.012 \\
2452008.918511 & 16.020 & 0.011 \\
2452008.920780 & 16.016 & 0.011 \\
2452011.918833 & 15.842 & 0.014 \\
2452011.923358 & 15.854 & 0.013 \\
2452012.059320 & 15.850 & 0.012 \\
2452012.061589 & 15.858 & 0.011 \\
2452012.921398 & 15.899 & 0.021 \\
2452012.923678 & 15.900 & 0.018 \\
2452012.925912 & 15.897 & 0.020 \\
2452013.054211 & 15.864 & 0.010 \\
2452013.058748 & 15.864 & 0.020 \\
2452013.061167 & 15.859 & 0.014 \\
2452013.871358 & 15.719 & 0.011 \\
2452013.874413 & 15.719 & 0.011 \\
2452013.876670 & 15.734 & 0.010 \\
2452014.057147 & 15.739 & 0.011 \\
2452014.059404 & 15.733 & 0.010 \\
2452014.061672 & 15.736 & 0.011 \\
2452014.915092 & 15.783 & 0.012 \\
2452014.917349 & 15.789 & 0.013 \\
2452015.915701 & 15.624 & 0.011 \\
2452015.917970 & 15.622 & 0.011 \\
2452015.920227 & 15.624 & 0.011 \\
\enddata  
\end{deluxetable}

The analysis presented here incorporates observations from four
sources.  The majority of the observations are from the MACHO Project
photometric database.  The collection process has been described in
detail elsewhere \citep{al95} so we give only a brief description
here. The MACHO observations were made with the refurbished 1.27m
Great Melbourne Telescope at Mount Stromlo Observatory (MSO), near
Canberra, ACT, Australia.  It was equipped with a prime focus
reimager-corrector with an integral dichroic beamsplitter which gave a
0.5 sq. deg field of view in two passbands simultaneously: a 450-590
nm MACHO $V$ filter and a 590-780 nm MACHO $R$ filter.  These were
each sampled with a 2$\times$2 array of 2048$\times$2048 Loral CCDs
which were read out concurrently via two amplifiers per CCD in about
70 seconds.  The image scale was 0.63 arcsec per pixel.  Data
reduction was performed automatically by Sodophot, a derivative of
DoPhot \citep{dophot}.  MACHO photometry was then transformed into
Cousins $V$ and $R$ bands for further interpretation \citep{al99}.

The eclipsing nature of 81.8997.87 was first reported by the Optical
Gravitational Lensing Experiment (OGLE) project \cite{ogle} as OGLE
LMC-SC16 119952.  OGLE observations were taken on the 1.3 m Warsaw
telescope at Las Campanas Observatory, Chile, operated by the Carnegie
Institute of Washington.  Photometry is in the standard $BVI$ bands
with the majority of the observations in the $I$ band.  MACHO and OGLE
data for this system were previously published in \cite{myfirst}.

\begin{figure}
\epsscale{1.0}
\plotone{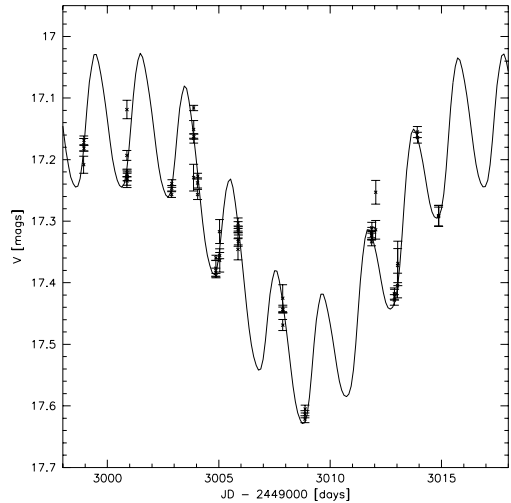}
\caption{$V$-band observations of the April 2001 primary eclipse
obtained on the 1.9 m telescope at Mount Stromlo Observatory along
with the curve of best fit.\label{oneecl-fig}}
\end{figure}

Follow-up observations of the April 2001 primary eclipse were obtained
in $BVI$ over 16 nights on the 1.9 m telescope at MSO.  It was
equipped with a 2048$\times$4096 SITe detector with a pixel size of 15
$\mu m^2$.  The chip was binned 5$\times$5 pixels resulting in a final
image scale of 0.45 arcsec per binned pixel.  Photometry was reduced
using the IRAF\footnote[1]{IRAF is distributed by the National Optical
Astronomy Observatories, which are operated by the Association of
Universities for Research in Astronomy, Inc., under cooperative
agreement with the National Science Foundation.}, DAOPHOT, ALLSTAR and
ALLFRAME packages.  On several nights images of an NGC 1866 standard
field were also taken and these observations were used to calibrate
the photometry to the NGC 1866 standards of \cite{walker}.  The
observations are tabulated in Tables \ref{B-tab}-\ref{I-tab} and the
$V$ observations can be seen in Figure \ref{oneecl-fig}.  The $B$ band
observations were not incorporated in the analysis presented here for
two reasons:
\begin{enumerate}
\item{The lack of $B$ photometry outside eclipse provides no useful
baseline for an analytical fit.}
\item{They provide very little insight into the nature of the
companion star as its color is so red (see below) that its
contribution to the $B$ flux is negligible.}
\end{enumerate}

\begin{deluxetable}{ccc}
\small
\tablecolumns{4} 
\tablewidth{0pt}   
\tablecaption{2MASS $JHK_s$ photometry of 81.8997.87 taken JD = 
2451580.5850.\label{2mass-tab}}
\tablehead{
\colhead{Filter}&
\colhead{Magnitude}&
\colhead{S/N}
}
\startdata 
$J$	& 14.421 $\pm$ 0.039	& 42.1\\
$H$	& 13.989 $\pm$ 0.050	& 28.7\\
$K_s$	& 13.606 $\pm$ 0.049	& 23.9\\
\enddata  
\end{deluxetable}

To complement the optical observations listed above, additional
photometry was extracted from 2MASS, a single-epoch all-sky survey in
the $JHK_s$ near-infrared bandpasses.  These data (Table
\ref{2mass-tab}) were obtained from the 2MASS all-sky point source
catalog, available online \citep{2mass}.

\section{Model and Results}

The model and fitting procedure used here is that described in detail
by \citet{myfirst} with only small modifications.  Most notably the
fit model has been adjusted to allow for eccentricity in the orbit
of the stars.  This is potentially a significant effect for this
system as the orbit of an 800-day binary is unlikely to have been
circularized.  However, given the poorly defined (or possibly poorly
covered) secondary eclipse in the current lightcurve its inclusion is
unlikely to produce a significant improvement in the fit.

\begin{figure}[hbtp]
\epsscale{1.0}
\plotone{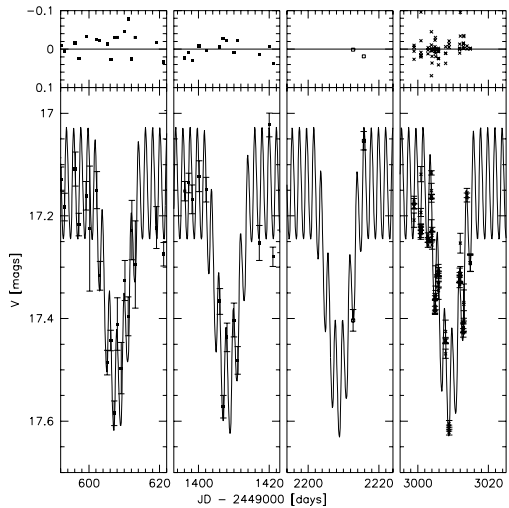}
\caption{Primary eclipses of 81.8997.87 in $V$ with the curve of best
fit.  The upper panels show residuals in magnitudes.  Filled boxes
indicate observations from the MACHO project, open boxes observations
from the OGLE project and crosses observations taken on the 1.9 m
telescope at Mount Stromlo Observatory.\label{pri-fig}}
\end{figure}

Figure \ref{pri-fig} shows the four primary eclipses of 81.8997.87 for
which we have observations along with the curve of best fit.  The
secondary eclipses are not shown because they cannot be clearly
separated from the Cepheid variations with the current observational
coverage (but they are included in the data published in
\cite{myfirst}).  The parameter set resulting from the fitting
procedure is shown in Table \ref{param-tab}.  The ratio of the surface
brightness in the $V$ and $R$ bands, $J_V/J_R$, a measurement of the
color, is tabulated as it was the property that was fit directly and
the other surface brightesses were computed from it.  Each $J_\lambda$
is expressed relative to the central surface brightness of the star.
For the Cepheid we give the mean $J_V/J_R$ to which a third order
Fourier series, representing the intrinsic temperature change of the
Cepheid, was added.  For the Cepheid we also tabulate $r_{min}$, the
star's minimum radius and $\Delta R_{amp}$, the amplitude of the
change in radius.  Both are expressed in units of the orbital
separation of the two stars as is the radius of the companion, $r$.
The Cepheid's pulsation period, $P_{Ceph}$, and $\Delta R_{shift}$,
the offset of the time of the Cepheid's minimum radius from the time
zeropoint of the data, are both given in days.  The uncertainties on
these parameters are determined from the covariance matrix of the
fitted parameters.  From the surface brightnesses, radii and magnitude
zeropoints, the intensity-weighted mean magnitude of each star in each
filter is calculated.  The mean colors are computed from the surface
brightness ratios described above, not the individual magnitude
values.  Also tabulated for the Cepheid is the value of $W_R =
R-3.0(V-R)$ where $3.0\sim A_R/(A_V-A_R)$ for this system (see below).
This index will correct for most of the effects of reddening and
differences in effective temperature between Cepheids.  The
uncertainty in the magnitude values are expressions of the range of
possible magnitudes based on the uncertainties in the fit parameters.
These are statistical uncertainties and likely underestimate the true
uncertainties in these parameters.  The orbital period is given in
days and the inclination, $i$, is in degrees.

\begin{deluxetable*}{lclc}
\tablecolumns{4}
\tablewidth{0pt}   
\tablecaption{Best-fit Parameters for 81.8997.87 ($\chi_\nu^2 = 1.5$)\label{param-tab}}
\tablehead{
	\multicolumn{2}{c}{Variable (Primary)}&
	\multicolumn{2}{c}{Companion (Secondary)}\\
	\hline
	\colhead{Parameter}&
	\colhead{Value}&
	\colhead{Parameter}&
	\colhead{Value}
}
\startdata
$\left<\frac{J_V}{J_R}\right>$& 1.046$\pm$0.008		&$\frac{J_V}{J_R}$	&0.51$\pm$0.08\\
$r_{min}$& 			0.0365$\pm$0.0011 	&$r$			&0.047$\pm$0.004\\
$\Delta R_{amp} $& 		0.0014$\pm$0.0002 	&...			&...\\
$P_{Ceph} (days)$&	 	2.035375$\pm$0.000009 	&...			&...\\
$\Delta R_{shift} (days)$& 	0.517$\pm$0.014 	&...			&...\\
\cutinhead{Intensity-weighted Mean Magnitudes and Colors}
$\left<V\right>$& 	17.2$\pm$0.2 	&$V$	&20.5$\pm$0.9\\
$\left<R\right>$& 	16.5$\pm$0.2 	&$R$	&19.0$\pm$0.9\\
$\left<I\right>$& 	15.9$\pm$0.2 	&$I$	&17.7$\pm$0.8\\
$\left<V-R\right>$& 	0.63$\pm$0.01 	&$V-R$	&1.4$\pm$0.3\\
$\left<V-I\right>$& 	1.325$\pm$0.005 &$V-I$	&2.72$\pm$0.07\\
$\left<W_R\right>$& 	14.4$\pm$0.2 	&...	&...\\
\cutinhead{Orbital Parameters}
$P_{orbital} (days)$& 	800.41$\pm$0.03	&	&\\
$i (deg.)$& 		86.4$\pm$0.3 	&	&
\enddata
\tablecomments{Meanings of individual parameters and units are explained in the text.}
\end{deluxetable*}

\section{Discussion}

\begin{figure}
\epsscale{1.0}
\plotone{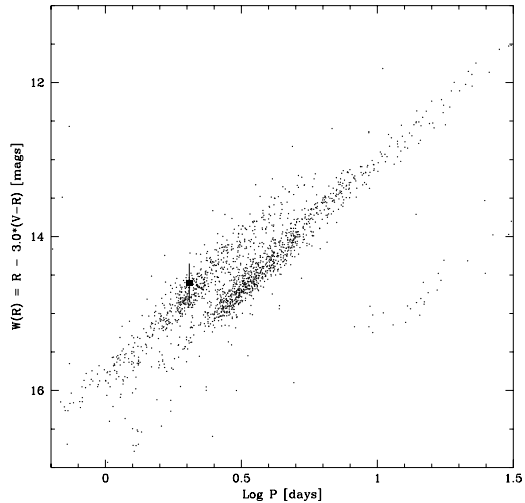}
\caption{$W_R$ vs. $\log P $ diagram for MACHO LMC Cepheids.  The
more luminous sequence at a given period contains Cepheids pulsating
in the first overtone mode and the sequence extending to longer
periods contains fundamental-mode pulsators.  The sequence in the
lower right contains the Type II (low-mass) Cepheids.  The location of
the variable component of this system (with the companion flux
removed) is indicated by the large black square. \label{pl-fig}}
\end{figure}

Figure \ref{pl-fig} shows that the location of the Cepheid in the
de-reddened Cepheid period-luminosity diagram is consistent with the
overtone Cepheid population.  This mode identification is confirmed by the
shape of the lightcurve measured through Fourier-fitting (with the
contamination due to the secondary removed).  At a period of $\sim$2
days the $R_{21}$ Fourier parameter cleanly distinguishes between
fundamental-mode and first-overtone Cepheids.  The small amplitude of
the change in radius ($0.041\pm0.001$ of the minimum Cepheid radius)
is also consistent with an overtone Cepheid.  The magnitudes and
colors in Table \ref{param-tab} are fainter and redder than expected
for a Cepheid, which we interpret as significant extinction along the
line of sight.  MACHO field 81 contains regions of considerable
star-formation activity.

\begin{figure}
\epsscale{1.0}
\plotone{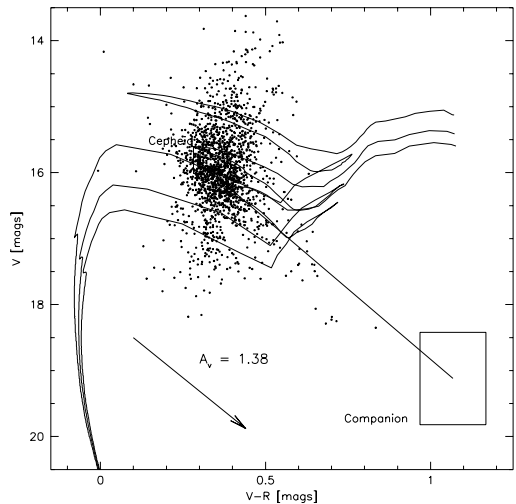}
\caption{Color-magnitude diagram for MACHO LMC Cepheids.  The boxes
indicate the error bounds on the properties of the two components
after the application of the extinction correction described in the
text.  The isochrones are from \protect{\cite{iso}} and represent
$\log_{10}(age)$ = 8.00, 8.14 and 8.25 (age in years).  The arrow is
the reddening vector for $A_V=1.38$ mag as derived in the text.
\label{cmd-fig}}
\end{figure}

By comparing the best-fit Cepheid magnitudes to those predicted by the
$V$ and $I$ period-magnitude relations for overtone Cepheids of
\cite{p-mag}, the amount of extinction in each bandpass and corrections
for the magnitudes of each star can be estimated.  These values are
found to be: $A_V = 1.38$ mag and $A_I = 0.67$ mag.  The relation
\begin{equation}
\left\langle \frac{A(I)}{A(V)} \right\rangle= 0.6800 - \frac{0.6239}{R_V}
\end{equation}
from \cite{ccm} yields $R_V = 3.15$ which, combined with the
corresponding relation for the $R$ band, gives $A_R = 1.04$.  Applying
these corrections we obtain the results shown in Figure \ref{cmd-fig}.
After correcting for extinction the companion's $V$ magnitude and
$V-R$ color seem consistent with a late-K or early-M class giant.

A possibility we considered is that the Cepheid is not an
intermediate-mass object but is instead a Type II Cepheid.  Mode
identification based on Fourier parameters is not well established for
Type II Cepheids of this period but based on the small photometric
amplitude (0.21 mags in $V$) this Cepheid would still be classified as
an overtone \citep{bl_her}.  The theoretical $P-L$ relation of
\cite{p2c_pl} gives $M_V = -1.06$ mags for a Type II overtone Cepheid
of this period, 1.78 magnitudes fainter than predicted for a Type I
overtone.  If zero extinction is assumed, $V=17.2$ mags given in Table
\ref{param-tab} implies a distance to the system of 44.8 kpc.  This
scenario is less likely because $V-R = 0.63$ mag for the Cepheid
implies a temperature that is too cool to be consistent with the
instability strip.  A significant amount of extinction (see above)
would need to be assumed with a concordant reduction in the assumed
distance.  This extinction could be Galactic or circumstellar.

\begin{figure}
\epsscale{1.0}
\plotone{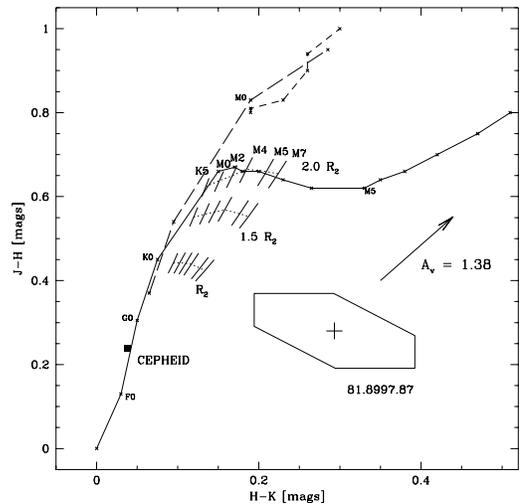}
\caption{Near-IR color-color diagram with fiducial sequences for main
sequence (solid black line) \protect{\citep{mstuff}}, giant (long-dashed black
line) and supergiant stars (short-dashed black line)\protect{\citep{allen}}.
The 2MASS colors of the system 81.8997.87 are shown by the black cross
and surrounding 1-$\sigma$ error box, both have been corrected for
extinction.  The large filled square indicates the theoretical colors
of a 2.035-day overtone Cepheid.  The three sequences show possible
combined colors of the two components assuming companion colors for
giant stars from \protect{\cite{mgiant}}.  The sequence labeled $R_2$ assumes
the two stars have radii in the ratio given by Table \protect{\ref{param-tab}}.
The sequences labeled $1.5~R_2$ and $2.0~R_2$ show the effect of
increasing the companion's radius by a factor of 1.5 and 2.0,
respectively.  The arrow is the reddening vector for $A_V=1.38$ mag as
derived in the text.\label{jhk-fig1}}
\end{figure}

The identification of the companion as a late-K or early-M class giant
does not seem to be borne out by the infrared observations.  The
observed 2MASS colors for 81.8997.87 (corrected for extinction) along
with the fiducial sequences for main sequence, giant and supergiant
stars in the near-infrared color-color diagram are shown in Figure
\ref{jhk-fig1}.  Also shown are the theoretical location of a 2.035
day overtone Cepheid based on the relations of \cite{mat} and
sequences representing the combination of the theoretical Cepheid
colors with a range of possible companion colors.  The combined colors
were computed assuming the ratio of radii given in Table
\ref{param-tab} (for the sequence labeled $R_2$) and assigning the
companion colors for late-type giants from \cite{mgiant}.  The dotted
line connects the different assumed spectral types while the solid
lines reflect the range in possible colors arising from the
uncertainty in the stellar radii.  It is possible that the true radius
of the companion is severely underestimated in the current fit to
the (primarily) optical lightcurve.  Thus two additional sequences,
calculated by assuming that the companion's radius is 1.5 and 2.0
times larger than the value in Table \ref{param-tab}, are also shown.
The sequences for typical giant colors do not extend to cool enough
temperatures to match the observed colors of this system.  Only by
assuming a much redder companion could we reproduce the observed
colors of the system.  This indicates that the system is unlikely to
consist solely of an overtone Cepheid and another normal star.

It is possible that the 2MASS magnitudes are in error or that their
uncertainties have been underestimated.  All the statistics provided
in the catalog indicate that these observations are reliable: high
signal-to-noise ratio, good quality psf fit ($\chi^2_\nu=$ 1.29, 0.92
and 0.97 for $J$, $H$ and $K$ respectively), source detected on all
available frames.  A more likely explanation would be the presence of a
cool, contaminating object within the same resolving element as the
target or the presence of hot circumstellar dust.

Figure \ref{cmd-fig} also shows that a system consisting of only these
two stars is a poor fit to the expectations from standard, single-star
evolutionary theory.  It is possible that the system is not a binary
but a hierarchical triple system (see discussion of V1334 Cyg by
\cite{triple}).  It is also possible that at some point in its history
one of the components has undergone an episode of mass loss.

The possibility that this is a non-hierarchical triple system has been
investigated by adding the presence of a third source of flux, not
participating in the eclipses, to the model and testing whether this
improves the fit to the observations.  This would also rule out the
possibility of an unrelated star along the line of sight.  It was
found that the quality of the fit improved slightly but, with a change
in $\chi^2_\nu$ of 0.015, not by a statistically significant margin.
Thus, this result neither supports nor excludes the presence of
additional sources of flux.

To attempt to further clarify the nature of this system we can
estimate the properties of the variable star from the known properties
of Cepheids and from its period and overtone classification.
\cite{pmr_fo} give the following canonical relation between period and
radius for first overtone Cepheids:
\begin{eqnarray}
\log R = 1.250(\pm0.005) + 0.755(\pm0.007) \log P,\\
~\sigma = 0.005 \nonumber
\end{eqnarray}
where $R$ has units of solar radii and $P$ has units of days.  This
gives a predicted radius for our Cepheid of $30.4 \pm 0.4 R_\odot$.
The ratio of stellar radii given by our fit then implies a companion
radius of $39 \pm 4 R_\odot$.  This radius is consistent with that of
a giant star but is poorly constrained likely because of the absence of
information from the system's secondary eclipses.  Our best fit value
of $r_1 = R_1/a$ gives an orbital separation of $a = 834 \pm 28
R_\odot = 3.9 \pm 0.1 AU$.

An estimate for the mass of the Cepheid can be obtained from the
canonical Period-Mass-Radius relation for first overtone pulsators
\begin{eqnarray}
\log M  = -2.776(\pm0.004) - 1.661(\pm0.140)\log P\\  +
2.682(\pm0.185)\log R,~~~\sigma = 0.004 \nonumber
\end{eqnarray}
with $M$ in units of solar masses, $P$ in units of days and $R$ in
units of solar radii, taken from \cite{pmr}.  With our period and
radius values, this yields a Cepheid mass of $4.9 \pm 0.2 M_\odot$
which, for an 800.4 day orbital period and $a = 3.9 \pm 0.1 AU$, puts
the companion's mass at $7.3 \pm 1.4 M_\odot$.

The binary system described above would have maximum radial velocities
of $v_r\sin i = 31 \pm 4~km~s^{-1}$ for the primary and $v_r\sin i =
21\pm 4 ~ km~s^{-1}$ for the secondary.  Therefore observations of the
radial velocity curves would certainly be worthwhile provided a
precision of $\pm1~km~s^{-1}$ could be obtained.  

\begin{deluxetable*}{llll}
\tablecolumns{4}
\tablewidth{0pt}
\tablecaption{Predicted Dates of Future Eclipses \label{future-tab}}
\tablehead{
	\multicolumn{2}{c}{Primary Eclipse}&
	\multicolumn{2}{c}{Secondary Eclipse}\\
	\colhead{JD}&
	\colhead{UT}&
	\colhead{JD}&
	\colhead{UT}
}
\startdata
...		&...			  &2,453,209.22 	&2004 Jul 22 5.34\\
2,453,609.84 	&2005 Aug 26 20.22        &2,454,009.63 	&2006 Sep 30 15.18\\
2,454,410.25 	&2007 Nov 5 6.06          &2,454,810.03 	&2008 Dec 9 0.78\\
2,455,210.66 	&2010 Jan 13 15.90        &2,455,610.44 	&2011 Feb 17 10.62\\
2,456,011.06 	&2012 Mar 24 1.50         &2,456,410.84 	&2013 Apr 27 20.22\\
2,456,811.47 	&2014 Jun 2 11.34         &2,457,211.25 	&2015 Jul 7 6.06\\
2,457,611.87 	&2016 Aug 10 20.94        &2,458,011.66 	&2017 Sep 14 15.90\\
2,458,412.28 	&2018 Oct 20 6.78         &2,458,812.06 	&2019 Nov 24 1.50\\
\enddata
\end{deluxetable*}

\section{Conclusions}

We have presented additional optical observations and the first
near-infrared photometry of this system.  Combined with previously
published optical data they support several conclusions:
\begin{enumerate}
\item{Based on the updated set of optical magnitudes, colors and
relative radii we can classify the components.  They are most
consistent with an intermediate-mass overtone Cepheid with a late K or
M-type giant companion.}
\item{This result is inconsistent with the expectations from
evolutionary theory.  The companion is too cool and dim for the system
to match theoretical isochrones.}
\item{In the near-infrared, a companion with cooler colors than
standard giant stars is needed to replicate the observed system
color.}
\end{enumerate}

Clearly, more observations are needed to fully realize the
considerable potential of this system.  In particular one of the
principal sources of the uncertainty in the companion's properties is
the lack of observations of a secondary eclipse.  To facilitate
follow-up work Table \ref{future-tab} presents a table of predicted
future dates of primary and secondary eclipses.

Given the low temperature of the companion, observations taken at
near-infrared wavelengths should put a stronger constraint on the
companion's properties.  In particular, precise photometry taken
during the primary and secondary eclipses would allow better estimates
of the individual colors of each component so their location in Figure
\ref{jhk-fig1} would be better determined.  Observations taken on an
8m class telescope would have sufficient resolution to identify
possible sources of contamination within the crowded field.

The companion and Cepheid would appear to have similar fluxes between
$J~(1.22~\mu m)$ and $H~(1.63~\mu m)$ and therefore (given the
estimated radial velocities given above) radial velocity work should
be attempted with a high-resolution near-infrared spectrograph on an
8m-class telescope.  Near-infrared spectra could also provide a more
definitive classification of the companion star.

\acknowledgments{ 

The authors would like to acknowledge both the support given to this
project by the staff of Mt Stromlo Observatory and their travails in
the wake of the catastrophic fire of January 18, 2003.  We would also
like to thank Brian Cook and John Howard of the Canberra Astronomical
Society for obtaining observations on the MSO 1.9 m on three nights
for this programme.

We acknowledge research support from the Natural Sciences and
Engineering Research Council of Canada (NSERC) in the form of a
research grant to D.L.W.  These observations were obtained by
D.L.W. during a research leave at the Institute of Geophysics and
Planetary Physics (IGPP) at the Lawrence Livermore National Laboratory
and he acknowledges the support provided by Dr. K.H. Cook at that
time.

This publication makes use of data products from the Two Micron All
Sky Survey, which is a joint project of the University of
Massachusetts and the Infrared Processing and Analysis
Center/California Institute of Technology, funded by the National
Aeronautics and Space Administration and the National Science
Foundation.
}

\end{document}